\documentclass{article}

\usepackage[affil-it]{authblk}

\usepackage[margin=1in]{geometry}

\usepackage{graphicx}
\usepackage{subfigure}

\usepackage{wrapfig}

\usepackage{array}
\usepackage{multirow,booktabs}
\newcolumntype{L}[1]{>{\raggedright\let\newline\\\arraybackslash\hspace{0pt}}m{#1}}
\newcolumntype{C}[1]{>{\centering\let\newline\\\arraybackslash\hspace{0pt}}m{#1}}
\newcolumntype{R}[1]{>{\raggedleft\let\newline\\\arraybackslash\hspace{0pt}}m{#1}}

\usepackage{hyperref}       
\usepackage{url}            
\usepackage{booktabs}       
\usepackage{amsfonts}       
\usepackage{nicefrac}       
\usepackage{microtype}      

\newcommand{\pt}{\ensuremath{p_T}}

\title{Long Short-Term Memory (LSTM) networks with jet constituents for boosted top tagging at the LHC}

\author[1,2]{Shannon Egan}
\author[1]{Wojciech Fedorko}
\author[1]{Alison Lister}
\author[3]{Jannicke Pearkes}
\author[1]{Colin Gay}
\affil[1]{The University of British Columbia}
\affil[2]{McGill University}
\affil[3]{Stanford University}
\date{} 
\begin{document}

\maketitle

\begin{abstract}
  Multivariate techniques based on engineered features have found wide adoption in the identification of jets resulting from hadronic top decays at the Large Hadron Collider (LHC). Recent Deep Learning developments in this area include the treatment of the calorimeter activation as an image or supplying a list of jet constituent momenta to a fully connected network.  This latter approach lends itself well to the use of Recurrent Neural Networks. In this work the applicability of architectures incorporating Long Short-Term Memory (LSTM) networks is explored. Several network architectures, methods of ordering of jet constituents, and input pre-processing are studied. The best performing LSTM network achieves a background rejection of 100 for 50\% signal efficiency.  This represents more than a factor of two improvement over a fully connected Deep Neural Network (DNN) trained on similar types of inputs.
\end{abstract}

\section{Introduction} \label{sec:intro}
At the LHC, the importance of top tagging - the discrimination of jets originating from hadronic decays of the top quark from light-flavour and gluon originated jets - is increasing as the searches for physics beyond the Standard Model of particle physics~\cite{ttbar_res_ATLAS_13TeV, ttbar_res_ATLAS_8TeV, ttbar_res_CMS_8TeV, vlq_htx_CMS, vlq_single_zt_CMS, vlq_wbx_ATLAS, excited_b_wt_CMS, stop_CMS_0l_pub,stop_ATLAS_1l_pub, stop_ATLAS_0l,stop_ATLAS_1l_conf} and Standard Model measurements~\cite{ttbar_diff_allhad_ATLAS_13TeV, ttbar_diff_allhad_ATLAS_8TeV} explore higher and higher momentum ranges. The relative performance of a number of techniques based on expert-engineered jet features have been studied in both the CMS~\cite{cms_tagging_perf} and ATLAS~\cite{atlas_tagging_perf} experiments, including a deep neural network trained on jet substructure features~\cite{atlas_dnn}. Recent developments in the application of machine learning and deep learning to the problem of top quark and boson tagging have focused on image based approaches~\cite{jet_image_1,jet_image_11,jet_image_2,irvine,deep_top}. Recursive neural networks were studied in the context of $W$ boson tagging in Ref.~\cite{kyle} where variable length sets of four-momenta are used as the input to the network. A fully connected deep neural network accepting a list of of jet constituent properties: the transverse momentum ($p_T$), pseudorapidity ($\eta$) and azimuth angle ($\phi$), developed in Ref.~\cite{jannickednn} achieved a rejection factor of light quark and gluon originated jets of 45 at 50\% signal efficiency for jets with $p_T$ between 600 and 2500 GeV. A set of transformations including a Lorentz rotation (preserving the invariant properties like jet mass) was shown to be crucial to achieving that performance. This tagging method will subsequently be referred to as the DNN tagger. Here an extension to this work is presented. Instead of a fixed-length list of constituent momenta the the variable length sets of $(p_T, \eta, \phi)$ triplets are sequentially fed into a Long Short-Term Memory (LSTM) network.

\section{Signal and background modelling}

The jet samples required for network training were generated using Monte Carlo (MC) simulation.  Both the signal, from hadronic top quark decays, and background, from gluon and light quark jets, were generated at leading order using \textsc{pythia} v8.219~\cite{pythia8}. All samples are generated at 13~TeV centre-of-mass energy.  
The signal samples consist of Sequential Standard Model $Z'$ boson~\cite{zppaper} production with masses ranging from 1400-6360 GeV in which the $Z'$ decay to $t\bar{t}$ pairs; only hadronic decays of the top quarks are allowed.  Cuts are applied on the invariant mass of the $t\bar{t}$ system and on the top quark $p_T$ to ensure that the pseudorapidity distribution of the top jets is approximately equivalent to that of the background jets.  Background events are generated as Quantum Chromodynamics (QCD) "dijet" processes, including gluon-gluon, quark-gluon and quark-quark scattering, with the \pt\ of outgoing partons ranging from 470 to 2790 GeV.  
A large number of "soft" QCD interactions, referred to as minimum bias, are also generated for the modelling of pileup~\cite{pileup} - i.e. of multiple $pp$ interactions within the same bunch crossing.  

The detector response is simulated using the \textsc{delphes} v3.4.0 suite~\cite{delphes} running with the default emulation of the CMS detector and particle flow event-reconstruction~\cite{cms_pflow_1,cms_pflow_2} - known as energy flow in \textsc{delphes}.    
Minimum bias events are overlaid onto signal and background events; a random number (drawn from a Poisson distribution) of $pp$ collisions are overlaid. Two pileup scenarios are studied.  The first scenario emulates LHC conditions at the end of 2016 with an average of 23 pileup interactions (referred to as LHC2016 pileup) while the second has an average pileup of 50, emulating the conditions expected at the end of LHC Run 2 (2018).

\section{Jet selection and sample preparation}

\textsc{delphes} energy flow objects resulting from event reconstruction were clustered into high radius jets using the anti-$k_T$ algorithm~\cite{antikt} implemented in FastJet~\cite{fastjet}.  The jet radius ($R$), a parameter of the clustering which determines the minimum distance between the centres of two jets, is set to $R=1$\footnote{\label{coordinates} In this article a cylindrical coordinate system is adopted with the $z$-axis along the beamline. Polar angle is $\theta$, azimuthal angle is $\phi$. Transverse momentum \pt\ is the component of particle momentum in the $x-y$ plane. Pseudorapidity $\eta$ is defined as $\eta=-\ln\left(\tan\left(\frac{\theta}{2}\right)\right)$. The jet clustering parameter $R$ is defined as $R=\sqrt{\Delta y^2 + \Delta\phi^2}$ where $y$ is rapidity.}. 

On some samples, a trimming procedure~\cite{trimming} is applied to reduce contributions from pileup and underlying event. This involves using the $k_T$ algorithm~\cite{kt} to re-cluster energy flow objects into "subjets" with $R=0.2$. Jet constituents belonging to subjets contributing less than 5\% of the jet \pt\ are removed. On the samples where trimming was not applied, the same reclustering procedure was run to identify subjets, however no cleaning threshold was applied.

Additional cuts were applied to constrain the set to jets with $600\ \leq p_T \leq 2500$~GeV and $\eta \leq 2.0$.  The remaining jets were sub-sampled to achieve a flat distribution in \pt. The signal and background have matching distributions in $\eta$ and no further sampling or weighting was applied. Subsampling is performed to prevent the network from learning the underlying \pt\ distribution as the discriminating feature, and to mitigate the degradation in performance at high \pt\ that is typical of jet taggers.

The final sample consists of approximately 7 million jets, split evenly between signal and background.  The full sample is divided into 3 subsets which play different roles in the network training.  The first 80\% of jets form the training set, 10\% are assigned as validation samples, the final 10\% forms the test set, on which performance metrics are evaluated. The jet order is shuffled after each epoch. In addition to this sample, an independent set of 11 million jets (again comprised of 50\% signal, 50\% background) is generated for the final evaluation of the trained network.

\section{Jet pre-processing}

Jets are preprocessed as in Ref.~\cite{jannickednn}. 
The transverse momenta of the jet constituents are scaled by $1/1700$ to bring all inputs to the scale of unity and prevent node saturation. Further jet preprocessing incorporates domain-specific knowledge about jet physics in an attempt to reduce the dimensionality of the latent space. Jets are translated in $\eta$ - $\phi$ space so that the leading (highest \pt) subjet is shifted to $(0,0)$, or equivalently the leading subjet is pointing along the $x$ coordinate. Next a Lorentz transformation is applied, designed to rotate the jet about the $x$ axis so that so that the second highest \pt\ subjet is located in the $x-y$ plane with negative $y$ coordinate. This transformation does not result in the loss of jet mass information. The final pre-processing step is an conditional reflection with respect to the $x-y$ plane, such that the average jet \pt\ has a positive $z$ component. Information from a maximum of 120 jet constituents is supplied to the network.

\section{Constituent sequence ordering} \label{sec:ordering}

\begin{wrapfigure}{r}{0.5\textwidth}
\centering
\includegraphics[width=0.5\textwidth]{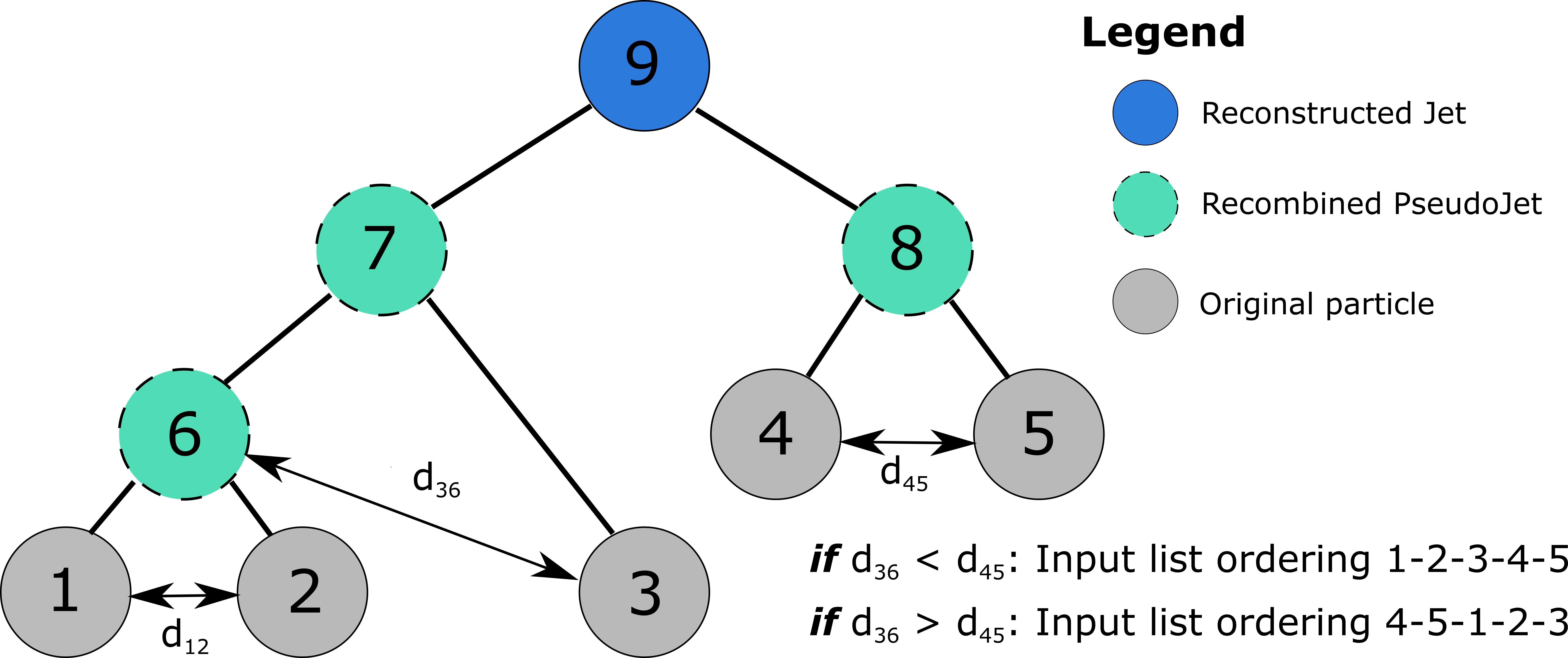}
\caption{An example of the binary tree constructed by jet algorithms during clustering and the resulting constituent list ordering presented to the LSTM in the substructure ordering scheme.}
\label{fig:tree}
\end{wrapfigure}

We hypothesize that the order of the constituent sequence can provide salient information for signal/background discrimination to the LSTM tagger, and thus develop sorting methods which attempt to represent the underlying QCD and substructure of the jets, referred to as substructure ordering.  In particular, we use a recursive algorithm which utilizes the history of the initial anti-$k_T$ clustering to add constituents to the input list in an order which reflects the jet substructure.  Clustering algorithms effectively produce a binary tree from the reconstructed particles, as depicted in Fig.~\ref{fig:tree}, where the intermediate jets are referred to as ``PseudoJets'' and are constructed by summing the four-momenta of the particles or PseudoJets with the smallest distance metric \footnote{The distance metric used is referred to as $d_{ij}$, $i$ and $j$ being indices of particles or PseudoJets in the event list, and is defined as: $d_{ij}=min(k^{2p}_{ti},k^{2p}_{ti})\frac{\Delta^2_{ij}}{R^2}$, where $k_{ti}$ is the transverse momentum of particle $i$. Exponent $p$ defines the precise algorithm used ($p=1$ for $k_T$, $p=-1$ for anti-$k_T$ or $p=0$ for Cambridge-Aachen), $R$ is the radius parameter of the clustering, and $\Delta_{ij}^2=(y_i-y_j)^2+(\phi_i-\phi_j)^2$, $y$ being the rapidity} at a given clustering step. The jet substructure sorting algorithm starts with the final jet and is called on each of the parent PseudoJets. Recursion is called on the pseudojet whose parents have a smaller $d_{ij}$. If one of the parents of the jet or PseudoJet under consideration is the original jet constituent that constituent is added to the list and recursion is continued on the other parent. If both parents of a jet or a PseudoJet are original constituents, both are added to the list with the higher \pt\ one added first and the recursion is terminated. Thus the ordering algorithm performs a depth-first traversal of the clustering tree.

This method is compared to sequence ordering schemes that were previously tested on the DNN in~\cite{jannickednn}, namely sorting purely by \pt\ of jet constituents, and ``subjet sorting''. In the latter scheme first subjets are arranged in a descending order by \pt, and then constituents of given subjet are added to the list, also in descending order by \pt. Subjet sorting was found to yield the best performance in~\cite{jannickednn}.

\section{Network architecture} \label{sec:archi}

The best-performing network design consists of an LSTM with state width of 128 connected to 64-node dense layer. Only the output of the LSTM layer at the last step is connected to the dense layer. This architecture was found through heuristic search of the number of LSTM layers, layer widths and presence or lack of the of the dense layer. Several optimization methods were tried with Adam~\cite{adam} providing the most stable training with highest final performance. The input data used for network selection was the trimmed, subjet sorted set with LHC2016 pileup. The Keras suite~\cite{keras} with the Theano~\cite{theano} backend was used to implement the model.

\section{Performance} \label{sec:performance}

\begin{figure}[t]
\centering
\subfigure{
\includegraphics[width=0.44\textwidth]{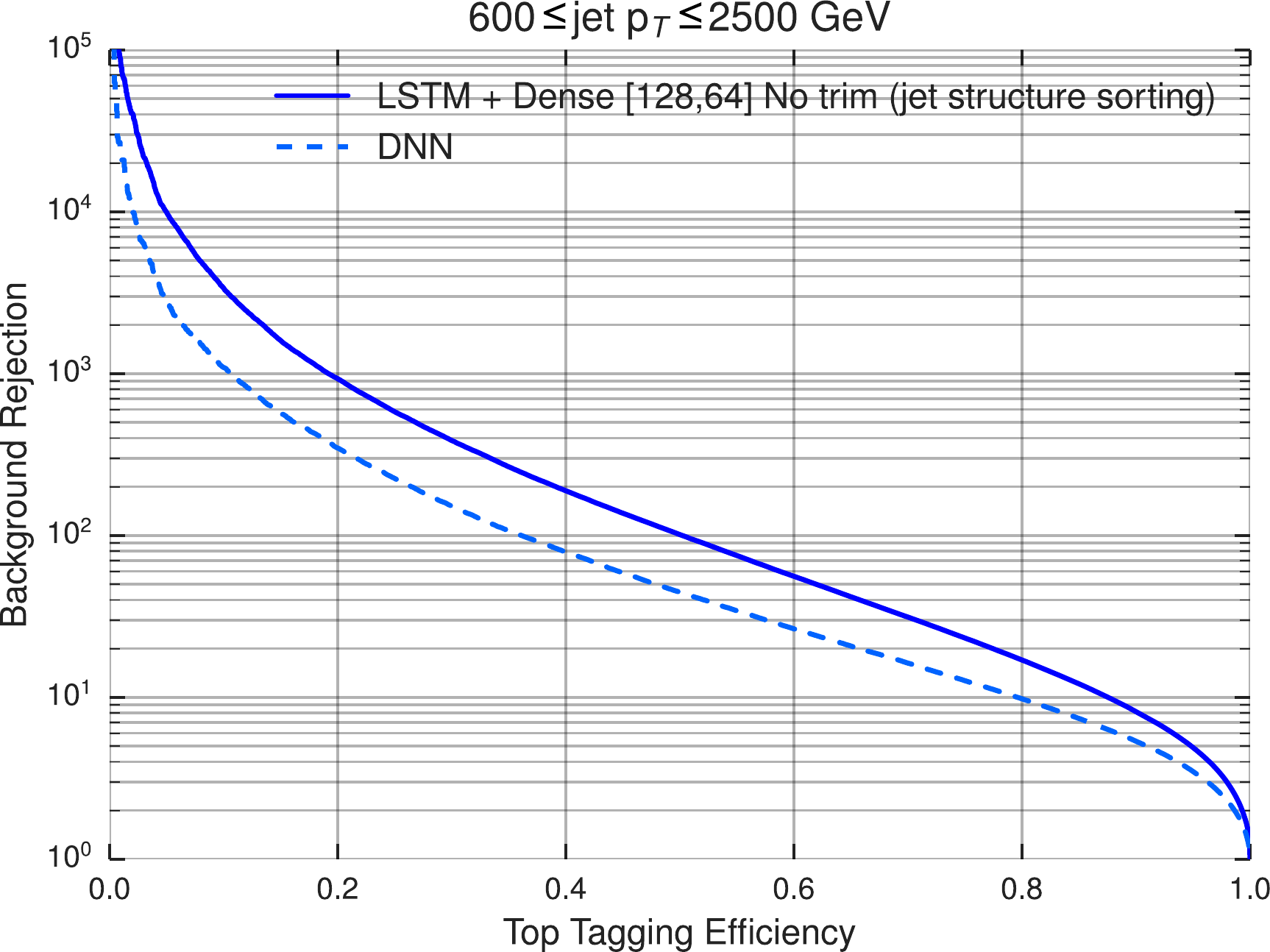}
}
\subfigure{
\includegraphics[width=0.44\textwidth]{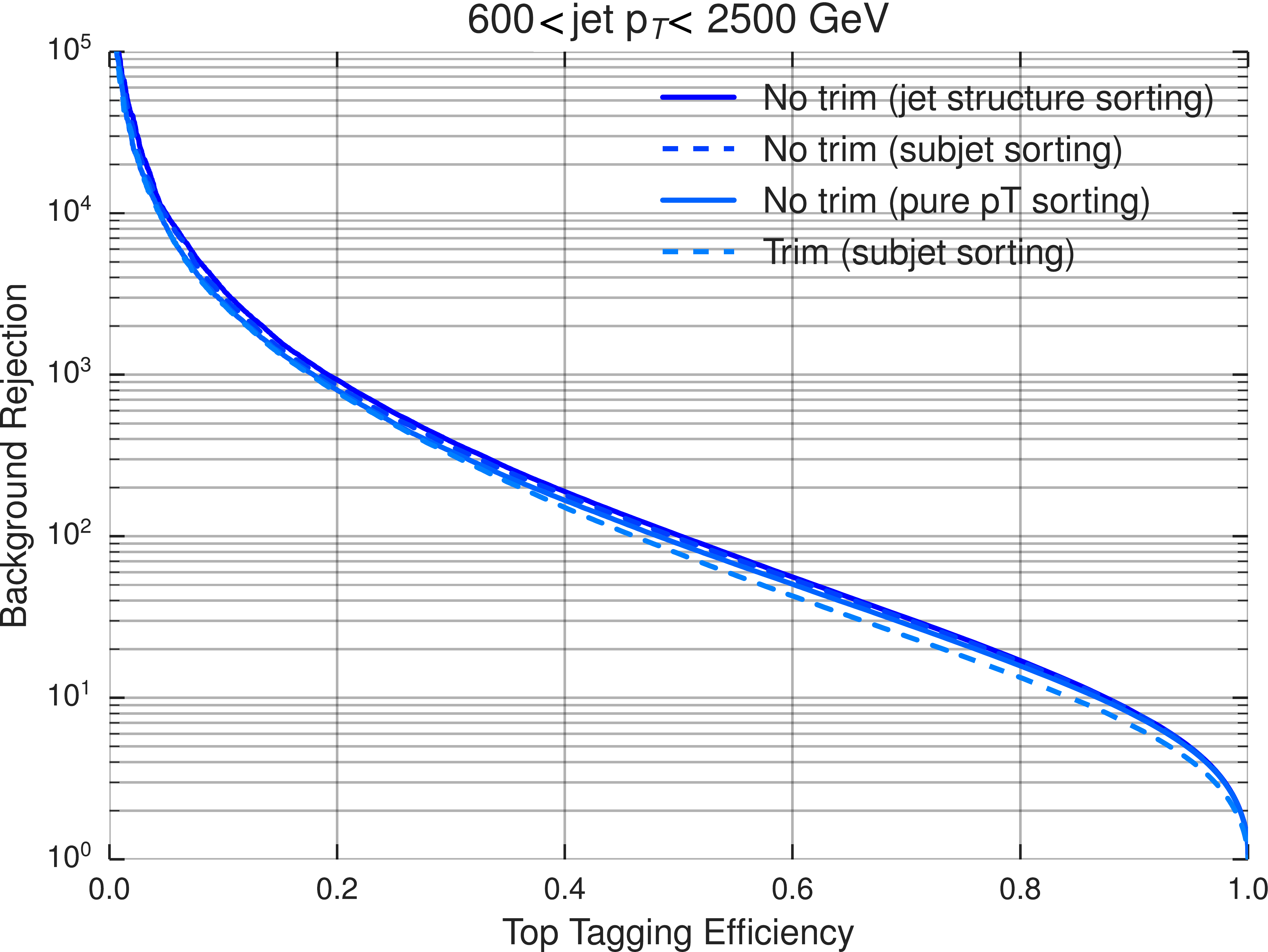}
}

\caption{Left panel shows the ROC curve comparing the best performing DNN tagger to best performing LSTM tagger under LHC 2016 pileup conditions.  Inputs to the DNN were trimmed and sorted by subjet, while LSTM inputs were untrimmed and sorted by the jet substructure-based method described in Section \ref{sec:ordering}. Right panel shows the ROC curve for best performing LSTM network under different trimming and constituent ordering conditions, under LHC 2016 pileup conditions. }
\label{fig:lstm_dnn}
\end{figure}

The primary interest of this study was to evaluate how an LSTM network would compare to the previously developed DNN.  Fig.~\ref{fig:lstm_dnn} (left) shows receiver operating characteristic (ROC) curves for the DNN and LSTM taggers under their respective best performing architectures and input conditions. The LSTM network yields better performance than the DNN across all signal efficiencies, in particular reaching a background rejection of 100 at 50\% signal efficiency - greater than a factor of two improvement with respect to the DNN. 

Table~\ref{tab:bkgrej} shows the background rejection power of the network when different pileup level datasets are analyzed and different constituent ordering schemes are used. The LSTM with substructure ordering displays a higher dependence on pileup conditions than the LSTM with subjet ordering, which has the best performance in high pileup conditions when not using trimming. This suggests that large pileup affects the jet clustering order. The full ROC curves for some of these combinations are shown in Fig.~\ref{fig:lstm_dnn} (right).

\begin{table}[h]
\centering
\begin{tabular}{cllcC{2cm}C{2cm}C{2cm}}
\toprule
\multirow{2}{*}[-0.7ex]{Architecture} & \multicolumn{3}{c}{Input conditions} & \multicolumn{3}{c}{Background rejection at signal efficiency of}\\
\cmidrule(r){2-4}
\cmidrule(r){5-7}
 & Pileup & Trim & Sorting &80\% & 50\% & 20\% \\
\cmidrule(r){1-7}
DNN & LHC 2016 & Yes & Subjet  &9.8&45&365\\
\cmidrule(r){1-1}
\cmidrule(r){2-4}
\multirow{6}{*}[-0.7ex]{LSTM} & \multirow{3}{*}{LHC 2016} & Yes & Subjet & 13.4 & 78& 780\\
                      & &No&Substructure&17.0&101&930\\
                      & &No  & Subjet & 16.7 & 97 &  855\\
\cmidrule(r){2-4}
                      & \multirow{3}{*}{50} &Yes& Subjet & 13.5&78&780\\
                      & & No &Substructure&16.1&93&790\\
                      & & No &Subjet&16.6&96&890\\
\bottomrule

\end{tabular}
\caption{Background rejection factors of the best performing LSTM network architecture given different input types and sorting method. The performance of the DNN network from~\cite{jannickednn} is shown for comparison.}
\label{tab:bkgrej}
\end{table}

\section{Conclusions}

This work shows that using a simple and relatively narrow LSTM network with a fully-connected projection improves greatly on a DNN top tagger using the exact same jet constituent inputs in list form. The best performing LSTM reaches a background rejection of 100 at 50\% signal efficiency for jets with $600 \leq p_T \leq 2500$~GeV, representing more than a factor of two improvement over the previously developed DNN. A new constituent sequence ordering method has been developed. It has been demonstrated that information encoded in the sequence ordering itself can impact the performance of the tagger as well as its response to pileup.   

\bibliographystyle{JHEP}

\bibliography{IPP_report_references}

\end{document}